# Anti-Interference Diffractive Deep Neural Networks for Multi-Object Recognition

Zhiqi Huang[1,2,#], Yufei Liu[3,#], Nan Zhang[1,2,#,*], Zian Zhang[1,2], Qiming Liao[1,2], Cong He[1,2], Shendong Liu[1,2], Youhai Liu[4], Hongtao Wang[5], Xingdu Qiao[6], Joel K. W. Yang[5], Yan Zhang[3,*], Lingling Huang[1,2,*], Yongtian Wang[1,2]

[1]Beijing Engineering Research Center of Mixed Reality and Advanced Display, School of Optics and Photonics, Beijing Institute of Technology, Beijing, China.
[2]National Key Laboratory on Near-surface Detection, Beijing, China.
[3]Beijing Key Laboratory of Metamaterials and Devices, Key Laboratory of Terahertz Optoelectronics of Ministry of Education, Capital Normal University, Department of Physics, Capital Normal University, Beijing, China.
[4]Qiyuan Lab, Building 8, No. 55 Zique Road, Haidian District, Beijing, China.
[5]Engineering Product Development, Singapore University of Technology and Design, Singapore, Singapore.
[6]Department of Electrical and Systems Engineering, University of Pennsylvania, Philadelphia, PA 19104, USA.

[#] These authors contributed equally: Zhiqi Huang, Yufei Liu, Nan Zhang
[*] E-mail: nanzhang@bit.edu.cn; yzhang@cnu.edu.cn; huanglingling@bit.edu.cn

## Abstract

Optical neural networks (ONNs) are emerging as a promising neuromorphic computing paradigm for object recognition, offering unprecedented advantages in light-speed computation, ultra-low power consumption, and inherent parallelism. However, most of ONNs are only capable of performing simple object classification tasks. These tasks are typically constrained to single-object scenarios, which limits their practical applications in multi-object recognition tasks. Here, we propose an anti-interference diffractive deep neural network (AI D$^2$NN) that can accurately and robustly recognize targets in multi-object scenarios, including intra-class, inter-class, and dynamic interference. By employing different deep-learning-based training strategies for targets and interference, two transmissive diffractive layers form a physical network that maps the spatial information of targets all-optically into the power spectrum of the output light, while dispersing all interference as background noise. We demonstrate the effectiveness of this framework in classifying unknown handwritten digits under dynamic scenarios involving 40 categories of interference, achieving a simulated blind testing accuracy of 87.4% using terahertz waves. The presented framework can be physically scaled to operate at any electromagnetic wavelength by simply scaling the diffractive features in proportion to the wavelength range of interest. This work can greatly advance the practical application of ONNs in target recognition and pave the way for the development of real-time, high-throughput, low-power all-optical computing systems, which are expected to be applied to autonomous driving perception, precision medical diagnosis, and intelligent security monitoring.

**Introduction**

Deep learning techniques have proven to be effective in the classification and localization of objects in multiple scenarios[1]. However, with the increasing complexity of application scenarios, current object recognition technologies face numerous challenges, such as simultaneously detecting multiple objects[2,3,4], especially when they occlude or overlap with each other. It is desired to develop a robust model for multi-object detection and instance segmentation capable of handling complex scenarios[5,6,7]. Besides, certain applications require highly precise recognition of fast-moving objects[8,9,10], which requires deep-learning-based high frame-rate processing capability and temporal information modeling ability[11,12,13,14]. To meet the above requirements, tensor core processors for object recognition must deliver low latency, high throughput, and exceptional energy efficiency[15]. Traditional digital computers, however, face limitations in speed and energy due to Joule heating, electromagnetic crosstalk, and parasitic capacitance[15,16,17,18]. Photonic technologies offer unparalleled advantages in the development of artificial intelligence hardware, providing solutions to overcome the bottlenecks of electronic systems[19].

Optical neural networks (ONNs) represent a promising neuromorphic computing paradigm for object recognition, enabled by their unique advantages in light-speed computation, ultra-low power consumption, and inherent parallelism. As a key milestone in the advancement of photonic technology, ONNs are drawing increasing attention[20,21,22,23,24,25,26,27]. In recent years, fundamental breakthroughs across diverse domains have accelerated the development of ONNs. Firstly, multi-dimensional multiplexing and interweaving techniques[28], including wavelength[28,29,30,31,32,33], polarization[34,35,36], and orbital angular momentum[37,38], enable massively parallel information processing[39]. Meanwhile, energy-efficient materials and components, such as non-volatile phase-change material in a micro-ring resonator array[40,41,42,43], low-loss lithium niobate modulators[44,45] and hybrid optoelectronic chips eliminating optical-electrical conversion[46], pave the way for the development of ONNs with significantly reduced system-level power consumption. Most notably, novel ONN architectures and training strategies are emerging[47,48,49,50], such as integrated diffractive-interference hybrid design[51], distributed computing architecture[51,52], and fully forward mode training methods[53], further enhancing inference performance and expanding functional diversity and complexity.

However, current ONNs are primarily designed for classifying a single target[54,55], which is far from the multi-object scenarios commonly encountered in real-world applications. Although some studies have investigated multiplexing schemes to facilitate multi-object classification[56], these schemes impose strict constraints on the location, size, and category of objects, thereby limiting the flexibility and practicality of the network. To overcome these limitations, several hybrid optoelectronic architectures have been proposed for dynamic scenes[57, 58]. However, these architectures require an electronic neural network as a post-processing module to further process the low-dimensional features extracted by ONNs. The need for analog-to-digital conversion introduces latency and increases power consumption, thus compromising the intrinsic advantages of optical computing. For multi-object and dynamic target recognition tasks, current ONNs encounter two major challenges regarding theory and optical systems. The reliance on scalar diffraction theory limits the ONNs' ability to model complex wavefront interactions and interferences among multiple spatially overlapping targets, while the static nature of most optical systems lacks reconfigurability and make ONNs fail to adapt instantaneously to dynamic input conditions.

Here, an anti-interference diffractive deep neural network (AI $D^2NN$) consisting of two transmissive diffractive layers is proposed, as schematically illustrated in Fig. 1. We employ distinct deep-learning-based training strategies to distinguish target with interference, and eliminate the impact of undesired interference on target's recognition result. In this work, targets are defined as handwritten digits 0-5 from the Modified National Institute of Standards and Technology (MNIST) dataset. To enhance the network's robustness against diverse forms of interference, an extensive interference dataset is utilized for training. This dataset includes intra-class interference from other handwritten digits 6-9 from MNIST dataset, inter-class interference derived from the Fashion-MNIST and EMNIST datasets. We also introduce dynamic interference by combining all of the aforementioned categories (40 categories in total) without any constraints on object location or size. The trained network can apply to multi-object scenarios, accurately and robustly recognizing targets and achieving a simulated blind testing accuracy of 87.4%. Furthermore, silicon-based metasurfaces are fabricated to physically implement the proposed AI $D^2NN$. We establish a terahertz (THz) experimental platform with 0.85 THz as the incident light source. Using this setup, the AI $D^2NN$ achieves a blind testing accuracy of 86.7%, which is in good agreement with the numerical simulations.

The presented framework exhibits excellent scalability and can be scaled physically to operate across a broad wavelength range of electromagnetic waves, simply by scaling the diffractive features in proportion to the wavelength range of interest. In addition, it can be seamlessly integrated with optical multi-dimensional multiplexing technologies, enabling more flexible and higher-capacity parallel recognition tasks for multiple targets. Therefore, it can greatly advance the practical application of ONNs in target recognition and pave the way for the development of real-time, high-throughput, low-power all-optical computing systems, which are expected to be applied to autonomous driving perception, precision medical diagnosis, and intelligent security monitoring.

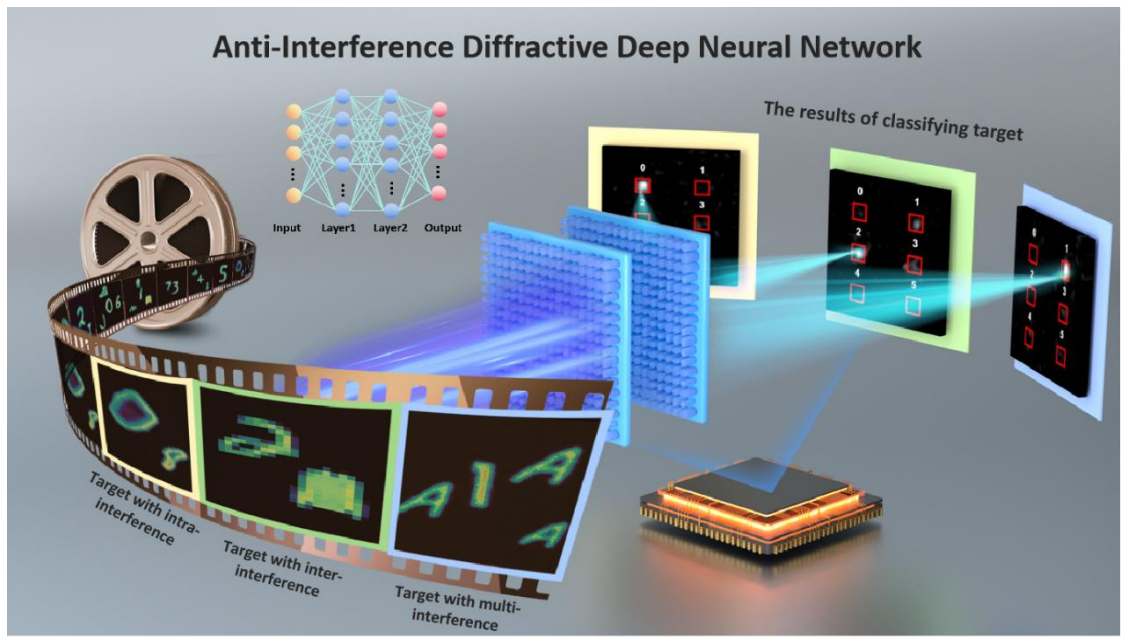

**Fig. 1 Schematic illustration of Anti-Interference Diffractive Deep Neural Network (AI D$^2$NN).** The network can classify handwritten digits (0-5) in multi-object scenarios, including intra-class interference, inter-class interference, and dynamic interference.

## Results

### Design of AI D$^2$NN

Current ONNs demonstrate reliable performance for single-object recognition or spatially constrained multi-object detection tasks, but exhibit substantial performance deterioration when applied to scenarios exceeding their functional scope, as shown in Fig. 2(a). The proposed AI D$^2$NN is capable of performing six-class classification of handwritten digits (0-5), even in the presence of undesired objects, as illustrated in Fig. 2(b). To achieve this, the basic training schematic is shown in Fig. 2(c). A customized training dataset was built by combining handwritten digits 0-5 with an interference set consisting of handwritten digits 6-9, fashions, and letters. Incorporating variability into the training dataset is intended to enhance the network's robustness in recognizing target objects, enabling selective identification even in complex scenarios with interference across categories, sizes, and positions.

Next, an all-optical neural network was constructed, which is composed of an input layer, multiple diffractive layers and an output layer. The diffractive layers function similarly to hidden layers in fully connected electronic neural networks, processing input information in a linear manner. Each unit within these layers performs complex amplitude modulation of the optical field, serving as a neuron node. Inspired by the Huygens-Fresnel diffraction principle, each neuron receives information in the form of a superposition of secondary optical fields generated by all neurons in the preceding layer, thereby tightly linking the layers and facilitating the hierarchical computation within the optical neural network. More detailed principal illustration of D$^2$NN can be found in Supplementary Note 1.

During training, the data was first fed into the network, and the output was obtained via optical diffraction calculations. The loss value was computed based on the loss function, as guidance through back-propagation for fine-tuning the weights of each hidden layer. This procedure highlights the importance of loss function design, and for this task, the loss function was specifically tailored according to different labels.

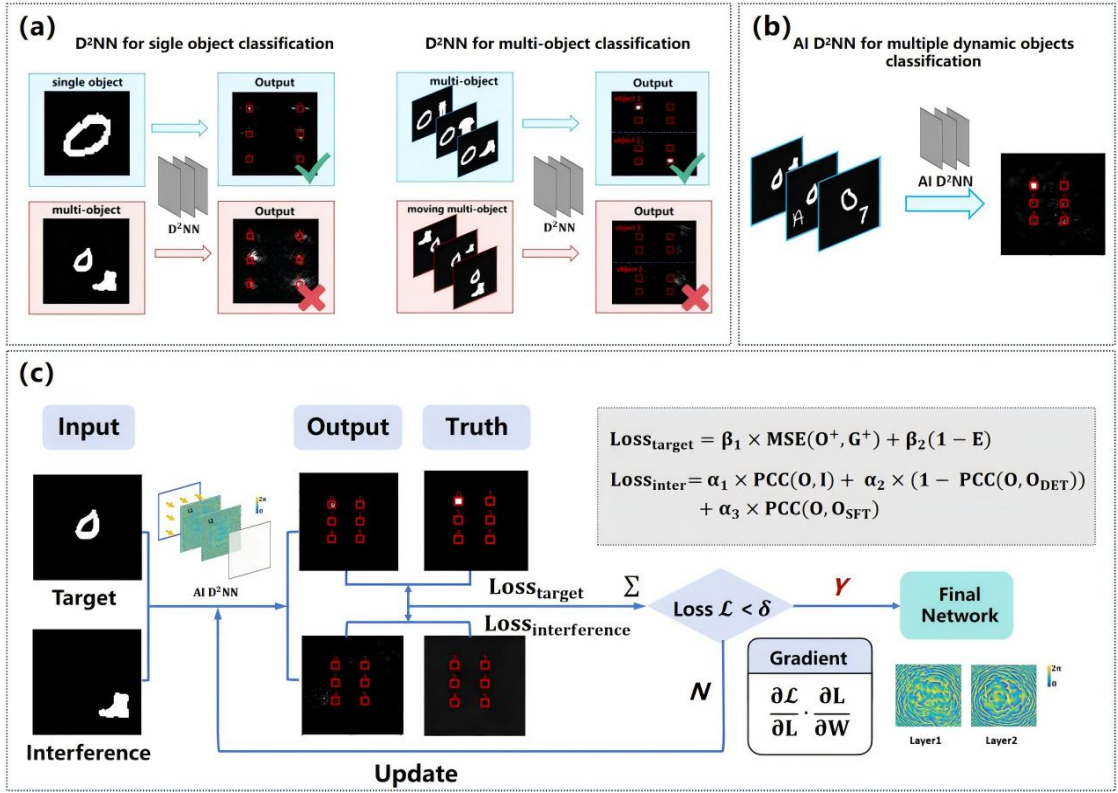

**Fig. 2 The principle of D²NN.** (a) Limitations of traditional D²NN when solving multi-object classification tasks and dynamic object classification tasks (b) The proposed AI D²NN is capable of performing six-class classification of handwritten digits (0-5) in the presence of multiple moving undesired objects (c) Training schematic of AI D²NN

For label 0-5, a conventional mean squared error (MSE) loss function was employed to maximize the light intensity in the classification region aligned with the target category and minimize the intensity in other classification regions. Moreover, constraints on the light field intensity was imposed to improve the energy efficiency E within the correct classification region.

$$Loss_{t\arg et} = \beta_1 \times MSE(O^+, G^+) + \beta_2 \times (1-E) \qquad (1)$$

For label 6, the output optical field of interference is uniformly distributed as random noise outside the classification regions, which was achieved by designing a loss function based on the Pearson Correlation Coefficient (PCC). PCC quantifies the degree of linear dependence between two variables, with values ranging from -1 to 1. An absolute value approaching 1 indicates a strong linear correlation, while a value of 0 signifies no linear relationship, which can be expressed mathematically as:

$$PCC(X,Y) = \frac{\text{cov}(X,Y)}{\sigma_X \sigma_Y} = \frac{\sum_{i=1}^{n}(X_i - \overline{X})(Y_i - \overline{Y})}{\sqrt{\sum_{i=1}^{n}(X_i - \overline{X})^2}\sqrt{\sum_{i=1}^{n}(Y_i - \overline{Y})^2}} \quad (2)$$

The loss function for label 6 was constructed as a superposition of three weighted PCC terms:

$$Loss_{inter} = a_1 \times PCC(O,I) + a_2 \times (1 - PCC(O,O_{DET})) + a_3 \times PCC(O,O_{SFT}) \quad (3)$$

Here, O and I means optical field distribution in output plane and input plane separately. $O_{DET}$ means optical field avoiding all classification regions. $PCC_{SFT}$ means optical field which shifts several pixels. PCC (O, I) aims to minimize the similarity between the input and output, preventing the direct transmission of interference information to the output. PCC (O, $O_{DET}$) guides the spatial distribution of the output optical field to bypass the detection region when this value trends toward 1, and the PCC (O, $O_{SFT}$) is minimized to force the $D^2NN$ to generate uninterpretable noise-like output patterns.

Within a batch, the sum of the loss values was calculated according to separate labels.

$$Loss = Loss_{target} + Loss_{inter} \quad (4)$$

The phase parameters of diffractive neurons were optimized using the backpropagation algorithm and stochastic gradient descent. After several epochs of training, the model converged and generated desired optical field distribution. More detailed information about training procedures and energy calculation methods are shown in Supplementary Note 2 and Supplementary Note 6.

To assess the classification accuracy of the network, a selective evaluation approach was designed. For digits 0-5, the classification was considered correct if the detection region with the maximum intensity corresponds to the correct label. For interference labeled 6, the recognition was deemed correct if both the mean and standard deviation of the optical output field were less than 0.2. This way provided an intuitive measure of the network's capability to suppress the influence of light field interference on digit classification results.

Additionally, other strategies were implemented to enhance the network's adaptability to experimental conditions. A 0.85 THz laser light source was employed for subsequent experiments. The impact of diffractive layer number, neuron sizes, and training iterations on network performance were systematically investigated. The diffractive surfaces, each comprising a 100×100 neuron array with respective neuron sizes of 100 μm and 200 μm, were compared under various experimental error conditions, including transverse shifts of 200 μm, rotational misalignments of 1 radian, and displacement errors along the z-axis of 100 μm, as shown in Fig. 3(a). Under ideal conditions, both networks exhibit comparable classification accuracies. However, the performance of these two networks degrades to different extents when introducing aforementioned experimental errors during training, as illustrated in Fig. 3(b)-(d). It can be concluded that the network with neuron sizes of 100 μm exhibits greater robustness compared to 200 μm counterpart. Considering fabrication constraints and alignment difficulty, 100 μm neuron size and a dual-layer configuration were selected for subsequent experiments. Subsequently, under the above network configuration, the effect of training epochs (from 1 to 10) on classification accuracy was further assessed, as shown in Fig. 3(e). The network achieves an accuracy exceeding 93% after just 4 epochs, and reached 93.7% after 10 epochs. More details about the combined effects of multiple experimental errors on network performance are available in Table S1 in Supplementary Note 3.

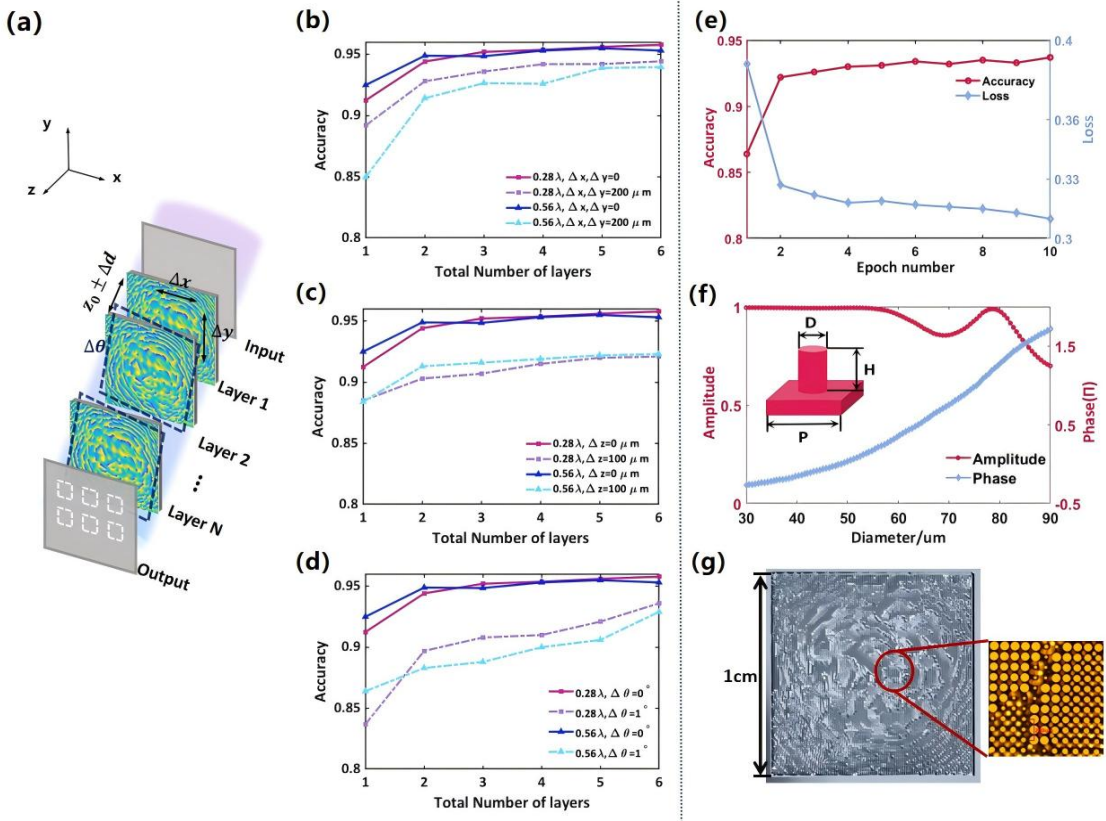

**Fig. 3 Training result of numerical network and optimization of microstructure parameters.** (a) Different error sources in the experiment (b)-(d) Comparison of training accuracy for networks with neuron sizes of 100 μm and 200 μm under interlayer transverse shift errors, z-axis displacement errors and rotational misalignment as the network depth increases (e) Training accuracy and loss of dual-layer AI D$^2$NN under 1 to 10 training epochs (f) Complex amplitude response of microstructure of different diameters under 0.85 THz (g) The overall distribution of the fabricated metasurface and zoom-in microscopic images of meta-atoms

**Design of the metasurfaces and elementary microstructures.**

As a proof of concept, the metasurface structures were designed for neuron phase modulation. The metasurface consists of high-impedance silicon cylinders deliberately fabricated on a silicon substrate. The electromagnetic response of each microstructure was calculated using the finite-difference time-domain (FDTD) method. The height of the microstructure was fixed at 240 μm, with a periodicity of 100 μm in both x and y directions. The incident wavelength was set at 0.85 THz.

To explore the phase modulation properties, the diameters of the nanostructures were varied from 30 μm to 90 μm in 300 nm increments. This yielded the complex response coefficients of the nanostructures under x-polarized incident light, as shown in Fig. 3(f). The 16 types of structures with amplitude coefficients close to 1 and phase responses nearest to integer multiples of Π/8 were selected. The specific structural parameters are listed in Table S2 in Supplementary Note 5. These structures were then assembled according to the neuron values of the trained diffractive neural network. Ultimately, two metasurfaces of 100×100 pixels were designed and fabricated for subsequent terahertz experiments. The overall distribution of the fabricated metasurface and the zoom-in optical microscopic images of metasurface are shown in Fig. 3(g).

**Anti-Interference Classification Results**

The proposed network is capable of accurately classifying the target in multi-object scenarios. Specifically, when the input contains both interference and target objects, the AI $D^2NN$ identifies the target, as evidenced by the concentration of energy within a designated region on the output plane. To evaluate the network's performance, both numerical simulations and experimental evaluations were conducted.

In the numerical simulations, a test dataset of 6,000 images containing digits and interference was built. These images were input into the trained network model to assess the classification accuracy of the primary digits. In the experiment, the input mask and two diffractive layers were assembled using 3D-printed holder, as shown in Fig. 4(a), completing the physical cascading of the AI $D^2NN$. A terahertz experimental platform was established, depicted in Fig. 4(b), and 60 samples were tested to evaluate the accuracy of the physical network. Detailed experimental set-up and fabricated samples are presented in Fig. S1 in Supplementary Note 4.

First, the network's classification capability in the presence of target digits 0-5 and intra-class interference digits 6-9 was evaluated. This case can mimic real-world industrial scenarios, such as automated inspection of specific types of components on a production line or the identification of vehicle types in transportation systems. The corresponding confusion matrix of the test dataset are shown in Fig. S3(a) in Supplementary Note 7, which indicates that the network achieved a recognition accuracy of 90.1% for digits 0-5 under scenes with intra-class interference. Fig. 4(c) demonstrates numerical simulation results and experimental results, which exhibit high consistency, proving that the delicately designed ONN metasurfaces possess the capability for target recognition.

Next, the network's classification capability was assessed when target and inter-class interference appear together. According to the confusion matrix in Fig. S3(b) in Supplementary Note 7, the network's recognition accuracy for digits 0-5 reaches 89.7% in the presence of inter-class interference. Both numerical simulation and experimental results are shown in Fig. 4(d). The energy distribution demonstrated that the system effectively concentrated most of the energy at the correct classification region. Although other classification regions appeared light spots, the maximum optical field intensity in these regions was only 50% of the intensity in the correct region. Furthermore, no significant light spots appeared outside the classification regions, indicating that the network can accurately classify target while remaining resistant to the influence of other interference present in the scene.

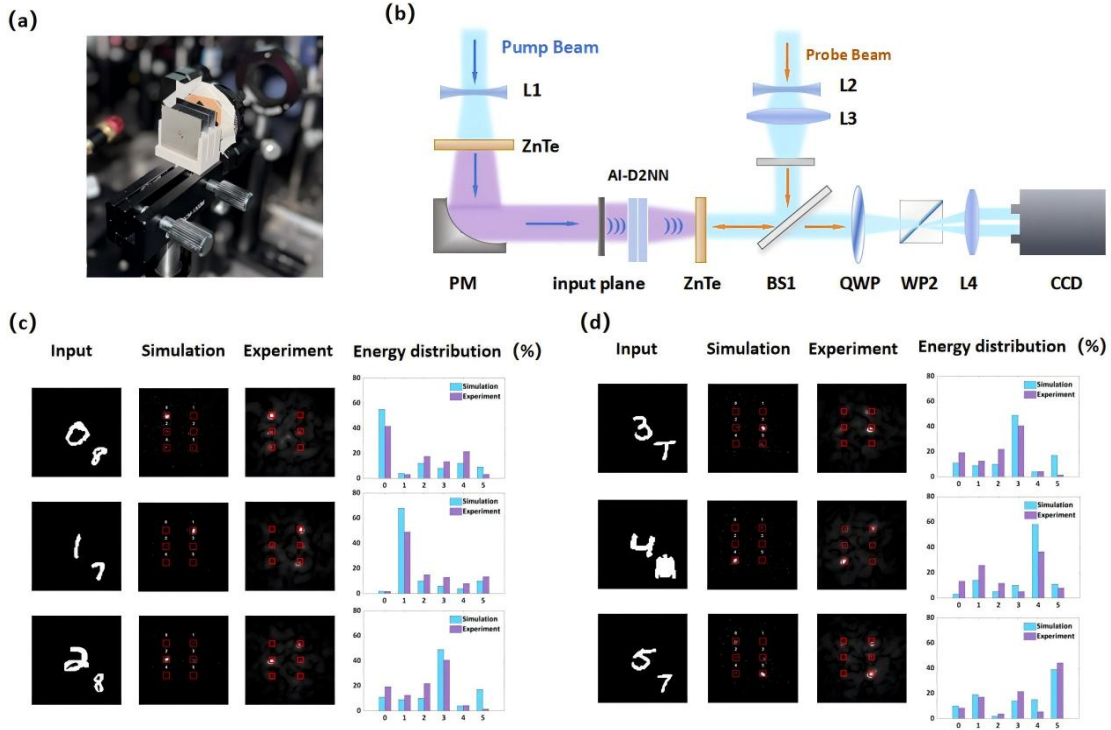

**Fig. 4 Anti-Interference Classification Results.** (a)-(b) Experimental ONN set up (c) The results of AI D$^2$NN classifying digits and intra-class interference (d) The results of AI D$^2$NN classifying digits and inter-class interference

To better simulate real-world scenarios, the dynamic multi-object scenes were constructed, where the target remains stationary while interfering objects allocate randomly in three-dimensional space. This configuration imposes natural challenges including objects clustering, partial occlusions, and scale variations, all arising organically from the objects' movement. During the dataset construction process, we simulated the movement of interfering objects parallel to the detection plane by varying their positions in the images, and simulated their movement perpendicular to the detection plane by scaling the interfering objects. By randomly altering the sizes and positions of the interfering objects, a series of dynamic multi-object scenarios was generated, as shown in Fig. 5(a). In numerical simulations, the network achieved an accuracy of 87.4% on a test set of 6,000 images. In Experimental validation, the classification accuracy reached 86.7% when each class was tested 10 times, closely matching the simulation results and confirming the robustness of the network design. The specific test results for examples of correct and incorrect identification are presented in Fig. 5(b). In addition, metrics such as signal-to-noise ratio (SNR) and the discrimination factor ($\triangle$E) were incorporated into the network evaluation metrics. In this study, SNR is defined as the ratio of the maximum energy among six classification regions to the total background energy outside these regions, while $\triangle$E represents the ratio of the difference between the highest and second-highest energies across the six detection regions to the highest energy. Numerical simulations yielded an SNR of approximately 31 and an $\triangle E$ value of approaching 1. Experimental measurements showed an SNR of approximately 23 and an $\triangle E$ value of 0.68, as presented in Fig. 5(c). Furthermore, Fig. 5(d) shows the network's output for the flexible test cases, demonstrating that the network can accurately classify the primary target regardless of the location and size of the interference. Besides, video streams were leveraged, which contains target (digit 4) and persistently moving interference as input. The corresponding output performance confirms the network's classification capability under dynamic scenes. The input and output videos are presented in Supplementary Video 1 and Fig. S5 in Supplementary Note 7.

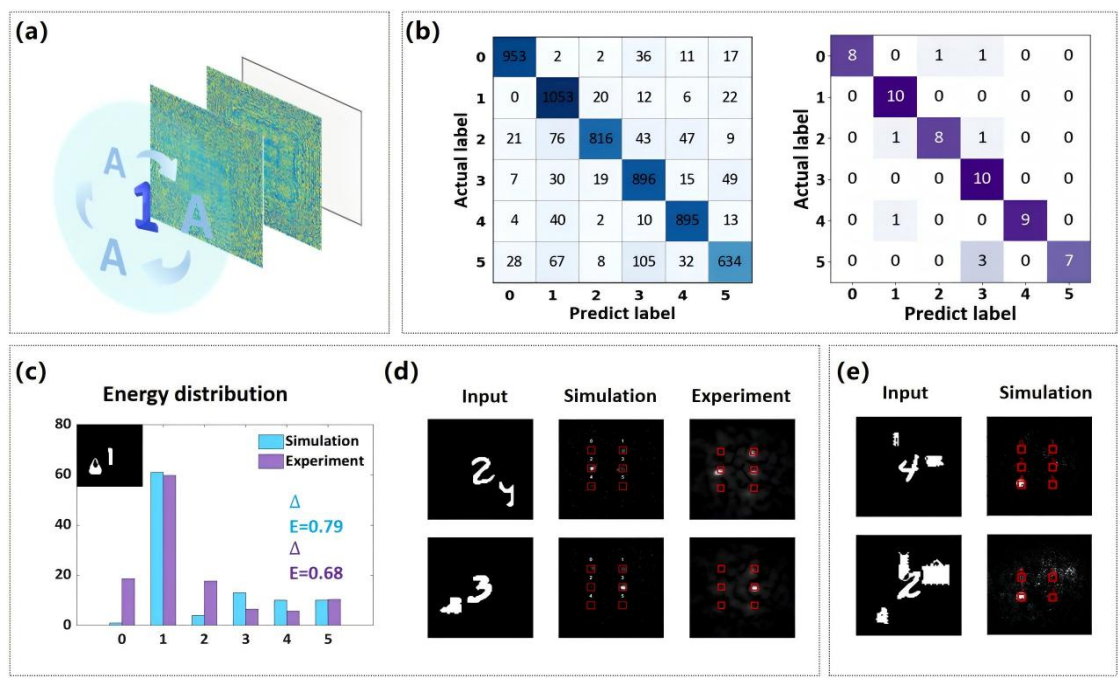

**Fig. 5 Anti-interference classification capability under dynamic multi-object scenes.**
(a) AI D$^2$NN under dynamic testing environment (b) Confusion matrices for the simulated results
(c) -(d) Energy distribution of numerical and experimental results when scenes containing one target and two moving interferences (e) Classification results when scenarios with one target and more than two moving interferences

The concept of transforming the optical fields of interfering objects into background noise through ONN is inherently extendable. Even as the number and categories of interference increase, the network still demonstrates excellent capability in classifying the primary target in multi-object scenarios. In theory, more complex scenarios for target recognition can be achieved by employing more diverse and extensive datasets, more rigorous loss functions and deeper network models during training phase. Therefore, we successfully constructed scenarios with one target and more than two interfering objects, in which the positions and sizes of all interfering objects vary flexibly. The model's target recognition accuracy and the output optical field distribution were evaluated using numerical simulations, as shown in Fig. 5(e) and Figs. S6-S8 in Supplementary Note 7, which demonstrated AI D$^2$NN has strong adaptability to complex scenarios. Lastly, a 50-frame video of the input and corresponding output was shown in Supplementary Video 2. The input scene contains a target (digit 2) and three interfering objects moving along randomly generated trajectories with varying sizes, further showcasing the network's scalability and generalization capability.

## Conclusion

Most existing ONNs are designed for single-object classification, and their performance deteriorates significantly when multiple objects present in a scene. To overcome this limitation, we proposed an anti-interference diffractive deep neural network for multi-object environments. Our approach enables the network to distinguish between target and interferences, and focus target's intensity onto designated detection region while transforming interferences' intensity into low-energy and uniformly distributed noise. This training strategy demonstrated strong robustness, achieving a numerical accuracy of 87.4% even in scenarios involving 6 types of targets and 40 types of dynamic interference. Furthermore, we constructed a terahertz-based diffractive neural network for experimental validation. The measured classification accuracy closely matched the simulation results, highlighting the network's potential for rapid and intelligent identification of primary targets in complex environments.

Future work can be developed regarding functional capability expansion, operational wavelength scaling and system-level integration. By integrating multi-dimensional optical encoding technology with AI D$^2$NN, each channel can be dedicated to recognizing only specific object classes while resisting interference from other objects, thus enabling multi-object recognition. In addition, by incorporating temporal features during feature extraction stage, AI D$^2$NN can process dynamic targets recognition. Besides, the operating wavelength of the network can be expanded to visible or near-infrared spectral range, to achieve broader applications. Furthermore, by capitalizing on the lightweight and planar advantages of metasurface, we can integrate metasurfaces with the mature complementary metal-oxide semiconductor imaging sensor, to fabricate an integrated and compact AI D$^2$NN system. Overall, AI D$^2$NN will accelerate their transition from laboratory prototypes to real-world applications, demonstrating strong potential for deployment in autonomous driving perception, precise medical diagnostics, and intelligent security monitoring.

## Materials and Methods

### 1. Numerical calculations

We employed the finite-difference time-domain (FDTD) method to simulate the optical response of microstructures with varying size. The modeled structures were subwavelength cylindrical rods made of high-impedance silicon (refractive index of 3.45), illuminated by a linearly x-polarized plane wave at 0.85 THz. The Si substrate thickness was set to 760 μm, and the height of each rod was fixed at 240 μm. The diameter of the nanostructures was varied from 30 μm to 90 μm in 300 nm increments. Periodic boundary conditions were applied in the x and y directions with a unit cell period of 100 μm to model an infinite array. Along the z-direction, a perfectly matched layer (PML) was employed to absorb outgoing waves and eliminate artificial reflections. To characterize the optical response, a point monitor was placed to record the transmitted wave phase, while a planar field positioned above the silicon rods captured the transmission intensity. By analyzing both phase and transmittance, a comprehensive insight into how rod's size interact with the incident terahertz waves were gained.

## 2. Training of the AI D$^2$NN

The numerical model of the proposed AI D$^2$NN was deployed on a computer for training. The hardware configuration includes an NVIDIA GeForce RTX 4060 GPU and an Intel® Xeon® Gold 6248R CPU. The software environment utilized Python 3.8.17 and PyTorch 2.0.1+cu117. During training, the mean squared error (MSE) loss function and Pearson correlation coefficient (PCC) was selected as the loss function, these metrics were commonly used for target classification and imaging quality assessment in machine learning tasks. The Adam optimizer was used to update the phase value of each layer in the network. We used a dataset consisting of 40,000 handwritten digit images and interference images for training, with a batch size of 8 and a learning rate of 0.01.

## 3. Experimental Setup

The AI D$^2$NN was experimentally implemented within a terahertz time-domain spectroscopy (THz-TDS) system. Femtosecond laser pulses were split into pump and probe paths: the pump beam generates terahertz radiation, while the probe beam was used to record the transmitted terahertz waveform after passing through the network. A CCD electro-optic sampling system was employed to scan the terahertz pulses, enabling the Fourier transformation of the recorded time-domain signals to obtain their spectral components in the frequency domain. To produce an ideal input intensity profile, a stainless-steel mask was used to block portions of the terahertz beam, creating a customized spatial intensity distribution. The custom 3D-printed holder ensured precise alignment and interlayer spacing between the mask and diffractive layers. Due to experimental uncertainties, cases such as slight transverse movement of the input mask relative to the metasurface occur. However, these misalignments only result in a slight shift of output pattern on the output plane, which don't affect final classification results.

# References

1. Kalake L., Wan W. G., Hou L. Analysis Based on Recent Deep Learning Approaches Applied in Real-Time Multi-Object Tracking: A Review. *IEEE Access* **9**, 32650-32671 (2021).
2. Sun S. J., Akhtar N., Song H. S., Mian A. S., Shah M. Deep Affinity Network for Multiple Object Tracking. *IEEE Transactions on Pattern Analysis and Machine Intelligence* **43**, 104-119 (2021).

3. Yang H., Zhou J. T., Cai J. F., Ong Y. S., IEEE. MIML-FCN plus : Multi-instance Multi-label Learning via Fully Convolutional Networks with Privileged Information. In: *30th IEEE/CVF Conference on Computer Vision and Pattern Recognition (CVPR)*) (2017).
4. Yang X. H.*, et al.* Deep Transfer Learning-Based Multi-Object Detection for Plant Stomata Phenotypic Traits Intelligent Recognition. *IEEE-ACM Transactions on Computational Biology and Bioinformatics* **20**, 321-329 (2023).
5. Chen Z. D., Ji H. B., Zhang Y. Q., Zhu Z. G., Li Y. F. High-Resolution Feature Pyramid Network for Small Object Detection on Drone View. *IEEE Transactions on Circuits and Systems for Video Technology* **34**, 475-489 (2024).
6. Zheng H. Y., Liu Q., Kravchenko, II, Zhang X. M., Huo Y. K., Valentine J. G. Multichannel meta-imagers for accelerating machine vision. *Nature Nanotechnology* **19**, 471-478 (2024).
7. Zaidi S. S. A., Ansari M. S., Aslam A., Kanwal N., Asghar M., Lee B. A survey of modern deep learning based object detection models. *Digital Signal Processing* **126**, 103514 (2022).
8. Kong L. Y., Yan Z. Y., Shi H. R., Zhang T., Wang L. LocaLock: Enhancing Multi-Object Tracking in Satellite Videos via Local Feature Matching. *Remote Sensing* **17**, 371 (2025).
9. Zhu T. Y.*, et al.* Looking Beyond Two Frames: End-to-End Multi-Object Tracking Using Spatial and Temporal Transformers. *IEEE Transactions on Pattern Analysis and Machine Intelligence* **45**, 12783-12797 (2023).
10. Wen T., Freris N. M. Semantically Enhanced Multi-Object Detection and Tracking for Autonomous Vehicles. *IEEE Transactions on Robotics* **39**, 4600-4615 (2023).
11. Chen T. I.*, et al.* Dual-Awareness Attention for Few-Shot Object Detection. *IEEE Transactions on Multimedia* **25**, 291-301 (2023).
12. Zhu L. C., Yang Y. Label Independent Memory for Semi-Supervised Few-Shot Video Classification. *IEEE Transactions on Pattern Analysis and Machine Intelligence* **44**, 273-285 (2022).
13. Wang L. M.*, et al.* Temporal Segment Networks for Action Recognition in Videos. *IEEE Transactions on Pattern Analysis and Machine Intelligence* **41**, 2740-2755 (2019).
14. Li Z. Y., Gavrilyuk K., Gavves E., Jain M., Snoek C. G. M. VideoLSTM convolves, attends and flows for action recognition. *Computer Vision and Image Understanding* **166**, 41-50 (2018).
15. Lin Z. J.*, et al.* 120 GOPS Photonic tensor core in thin-film lithium niobate for inference and in situ training. *Nature Communications* **15**, 1-10 (2024).
16. Feldmann J.*, et al.* Parallel convolutional processing using an integrated photonic tensor core. *Nature* **589**, 52-58 (2021).
17. Miller D. A. B. Attojoule Optoelectronics for Low-Energy Information Processing and Communications. *Journal of Lightwave Technology* **35**, 346-396 (2017).

18. Lee S. H., Zhu X. J., Lu W. D. Nanoscale resistive switching devices for memory and computing applications. *Nano Research* **13**, 1228-1243 (2020).
19. Feng F., *et al.* Symbiotic evolution of photonics and artificial intelligence: a comprehensive review. *Advanced Photonics* **7**, 024001 (2025).
20. Lin X., *et al.* All-optical machine learning using diffractive deep neural networks. *Science* **361**, 1004-1008 (2018).
21. Ashtiani F., Geers A. J., Aflatouni F. An on-chip photonic deep neural network for image classification. *Nature* **606**, 501-506 (2022).
22. Bai B. J., *et al.* All-optical image classification through unknown random diffusers using a single-pixel diffractive network. *Light-Science & Applications* **12**, 14 (2023).
23. Huang Z. B., *et al.* All-Optical Signal Processing of Vortex Beams with Diffractive Deep Neural Networks. *Physical Review Applied* **15**, 36936-36952 (2021).
24. Chen Z. J., *et al.* Deep learning with coherent VCSEL neural networks. *Nature Photonics* **17**, 723-730 (2023).
25. Hu J. T., Mengu D., Tzarouchis D. C., Edwards B., Engheta N., Ozcan A. Diffractive optical computing in free space. *Nature Communications* **15**, 1-21 (2024).
26. Wetzstein G., *et al.* Inference in artificial intelligence with deep optics and photonics. *Nature* **588**, 39-47 (2020).
27. Zou X. P. F., Huang X. W., Liu C., Tan W., Bai Y. F., Fu X. Q. Target recognition based on pre-processing in computational ghost imaging with deep learning. *Optics and Laser Technology* **167**, 1-9 (2023).
28. Jiang Y., Zhang W. J., Yang F., He Z. Y. Photonic Convolution Neural Network Based on Interleaved Time-Wavelength Modulation. *Journal of Lightwave Technology* **39**, 4592-4600 (2021).
29. Li J. X., Gan T. Y., Bai B. J., Luo Y., Jarrahi M., Ozcan A. Massively parallel universal linear transformations using a wavelength-multiplexed diffractive optical network. *Advanced Photonics* **5**, 016003 (2023).
30. Chi H. X., *et al.* Metasurface Enabled Multi-Target and Multi-Wavelength Diffraction Neural Networks. *Laser & Photonics Reviews* **19**, 158 (2025).
31. Duan Z. Y., Chen H., Lin X. Optical multi-task learning using multi-wavelength diffractive deep neural networks. *Nanophotonics* **12**, 893-903 (2023).
32. Xiang C. X., Qiu J. M., Liu Q. G., Xiao S. Y., Liu T. T. Multiplexed metasurfaces for diffractive optics via a phase correlation method. *Optics Letters* **50**, 1989-1992 (2025).
33. Luo Y., *et al.* Design of task-specific optical systems using broadband diffractive neural networks. *Light-Science & Applications* **8**, 112 (2019).
34. Li J. X., Hung Y. C., Kulce O., Mengu D., Ozcan A. Polarization multiplexed diffractive computing: all-optical implementation of a group of linear transformations through a polarization-encoded diffractive network. *Light-Science & Applications* **11**, 153 (2022).

35. Luo X. H., *et al.* Metasurface-enabled on-chip multiplexed diffractive neural networks in the visible. *Light-Science & Applications* **11**, 158 (2022).
36. Wang Y. Z., Yu A. X., Cheng Y. Y., Qi J. R. Matrix Diffractive Deep Neural Networks Merging Polarization into Meta-Devices. *Laser & Photonics Reviews* **18**, 2300903 (2024).
37. Li B. L., *et al.* Orbital angular momentum optical communications enhanced by artificial intelligence. *Journal of Optics* **24**, 094003 (2022).
38. Fang X. Y., *et al.* Orbital angular momentum-mediated machine learning for high-accuracy mode-feature encoding. *Light-Science & Applications* **13**, 49 (2024).
39. Cheng J. W., *et al.* Multimodal deep learning using on-chip diffractive optics with in situ training capability. *Nature Communications* **15**, 6189 (2024).
40. Wei M. L., *et al.* Electrically programmable phase-change photonic memory for optical neural networks with nanoseconds in situ training capability. *Advanced Photonics* **5**, 046004 (2023).
41. Xu S. F., Wang J., Yi S. C., Zou W. W. High-order tensor flow processing using integrated photonic circuits. *Nature Communications* **13**, 7970 (2022).
42. Tait A. N., *et al.* Neuromorphic photonic networks using silicon photonic weight banks. *Scientific Reports* **7**, 7430 (2017).
43. Wu C. M., Yu H. S., Lee S., Peng R. M., Takeuchi I., Li M. Programmable phase-change metasurfaces on waveguides for multimode photonic convolutional neural network. *Nature Communications* **12**, 96 (2021).
44. Guarino A., Poberaj G., Rezzonico D., Degl'Innocenti R., Günter P. Electro-optically tunable microring resonators in lithium niobate. *Nature Photonics* **1**, 407-410 (2007).
45. Zhu D., *et al.* Integrated photonics on thin-film lithium niobate. *Advances in Optics and Photonics* **13**, 242-352 (2021).
46. Chen Y. T., *et al.* All-analog photoelectronic chip for high-speed vision tasks. *Nature* **623**, 48-57 (2023).
47. He C., *et al.* Pluggable multitask diffractive neural networks based on cascaded metasurfaces. *Opto-Electronic Advances* **7**, 230005 (2024).
48. Bai B. J., *et al.* Pyramid diffractive optical networks for unidirectional image magnification and demagnification. *Light-Science & Applications* **13**, 178 (2024).
49. Bai B. J., *et al.* To image, or not to image: class-specific diffractive cameras with all-optical erasure of undesired objects. *eLight* **2**, 14 (2022).
50. Ma G. D., *et al.* Unidirectional imaging with partially coherent light. *Advanced Photonics Nexus* **3**, 066008 (2024).
51. Xu Z. H., Zhou T. K., Ma M. Z., Deng C. C., Dai Q. H., Fang L. Large-scale photonic chiplet Taichi empowers 160-TOPS/W artificial general intelligence. *Science* **384**, 202-209 (2024).
52. Zhang H., *et al.* An optical neural chip for implementing complex-valued neural network. *Nature Communications* **12**, 457 (2021).

53. Xue Z. W., Zhou T. K., Xu Z. H., Yu S. L., Dai Q. H., Fang L. Fully forward mode training for optical neural networks. *Nature* **632**, 280-286 (2024).
54. Li J. X., Mengu D., Luo Y., Rivenson Y., Ozcan A. Class-specific differential detection in diffractive optical neural networks improves inference accuracy. *Advanced Photonics* **1**, 046001 (2019).
55. Zhu H. H.*, et al.* Space-efficient optical computing with an integrated chip diffractive neural network. *Nature Communications* **13**, 1044 (2022).
56. Chi H. X.*, et al.* Metasurface Enabled Multi-Target and Multi-Wavelength Diffraction Neural Networks. *Laser & Photonics Reviews* **19**, 2401178 (2025).
57. Qian C.*, et al.* Dynamic recognition and mirage using neuro-metamaterials. *Nature Communications* **13**, 2694 (2022).
58. Huang Z., Shi W. X., Wu S. K., Wang Y. D., Yang S. G., Chen H. W. Pre-sensor computing with compact multilayer optical neural network. *Science Advances* **10**, eado8516 (2024).

# Supplementary information for

# Anti-Interference Diffractive Deep Neural Networks for Multi-Object Recognition

Zhiqi Huang[1,2,#], Yufei Liu[3,#], Nan Zhang[1,2,#,*], Zian Zhang[1,2], Qiming Liao[1,2], Cong He[1,2], Shendong Liu[1,2], Youhai Liu[4], Hongtao Wang[5], Xingdu Qiao[6], Joel K. W. Yang[5], Yan Zhang[3,*], Lingling Huang[1,2,*], Yongtian Wang[1,2]

[1]Beijing Engineering Research Center of Mixed Reality and Advanced Display, School of Optics and Photonics, Beijing Institute of Technology, Beijing, China.
[2]National Key Laboratory on Near-surface Detection, Beijing, China.
[3]Beijing Key Laboratory of Metamaterials and Devices, Key Laboratory of Terahertz Optoelectronics of Ministry of Education, Capital Normal University, Department of Physics, Capital Normal University, Beijing, China.
[4]Qiyuan Lab, Building 8, No. 55 Zique Road, Haidian District, Beijing, China.
[5]Engineering Product Development, Singapore University of Technology and Design, Singapore, Singapore.
[6]Department of Electrical and Systems Engineering, University of Pennsylvania, Philadelphia, PA 19104, USA.

[#] These authors contributed equally: Zhiqi Huang, Yufei Liu, Nan Zhang
[*] E-mail: nanzhang@bit.edu.cn; yzhang@cnu.edu.cn; huanglingling@bit.edu.cn

This file includes:
Supplementary Note 1: Angular spectrum diffraction method
Supplementary Note 2: The training of AI D$^2$NN
Supplementary Note 3: Analysis of robust training
Supplementary Note 4: Detailed experimental setup and fabricated samples
Supplementary Note 5: The structural parameters of cylindrical rods
Supplementary Note 6: Energy distributions for the AI D$^2$NN
Supplementary Note 7: Trained network's classification results of different cases

**Supplementary Note 1: Angular spectrum diffraction method**

In AI D²NN, each meta-unit behaves like a separate neuron in a neural network, interconnecting with the meta-units of the previous and next layers by diffraction of light. This physical process can be described by a mathematical model through the angular spectrum method.

Assume the complex amplitude distribution of input plane is U(x,y,0), field distribution at distance z is U(x,y,z), which can be expressed using Angular spectrum diffraction method.

$$U(x,y,z) = F^{-1}\{F[U(x,y,0)] \cdot H(f_x, f_y, z)\} \tag{S1}$$

Here, F and $F^{-1}$ denote the Fourier transform and inverse Fourier transform, respectively, and H represents the transfer function. The angular spectrum method decomposes the complex amplitude distribution on the input plane into plane waves of different spatial frequencies through the Fourier transform. Each component's optical field after propagation along the axial distance z is calculated separately, and the spatial distribution of the optical field at distance z is obtained through the inverse Fourier transform. In this process, the transfer function can be expressed as:

$$H(f_x, f_y, z) = e^{ik_z z} \tag{S2}$$

Here, $K_z$ is the wave vector component along the propagation direction. This transfer function describes the change of complex amplitude as a plane wave propagates along the z-axis. Meanwhile, the optical field is further modulated by the diffractive layer neurons during propagation, which is represented as the product of the complex amplitude transmission coefficient and the optical field, the mathematical expression is as follows:

$$u_n^{l+1}(x,y) = \iint_s F^{-1}\{F\{u_n^l(x_n, y_n) t_n^l(x_n, y_n)\} H(f_x, f_y, z)\} \tag{S3}$$

Therefore, the input optical field at node n of the next layer l+1 is obtained by receiving and summing up all the secondary optical fields generated by the neurons in the previous layer l. Repeat the process described by formula (S3) to implement multi-layer diffractive neural network. It should be noted that $t_n^l(x_n, y_n)$ represents the complex - amplitude modulation of the optical field by the meta-unit, since we select nanopillar structures with a transmittance close to 1, $t_n^l(x_n, y_n)$ can be simplified to

$\varphi_n^l(x_n, y_n)$, and subsequent optimizations are also performed specifically for $\varphi_n^l(x_n, y_n)$.

**Supplementary Note 2: The training of AI D²NN**

This section provides further details to supplement the discussion of network principles presented in the main text. To enable optical neural networks (ONNs) to achieve target classification in multi-object scenarios, we focus on specific designs in **datasets construction and loss function configuration.**

**(1) Dataset**

We customized a train dataset which is composed of a target dataset and interference dataset. The target dataset consists of six types of handwritten digits (0-5) from MNIST, while interference dataset includes other handwritten digits of MNIST (6-9), 10 types of fashions in Fashion MNIST dataset, and 26 types of letters in EMNIST dataset. The processing of the dataset images is divided into the following steps:

1. Binarization: Converting images into binary form according to a preset binary threshold highlights the object's contour information, ultimately improving the model's recognition accuracy.

2. Scaling: Images in the interference dataset are resized from $10 \times 10$ pixels to $28 \times 28$ pixels.

3. New target dataset for training: Each image from the target dataset (28×28 pixels) is placed in the center of an all-zero image with a resolution of $56 \times 56$ pixels. A total of 20,000 images are generated, with labels directly corresponding to the digits (0-5).

4. New interference dataset for training: Each image from the interference dataset is placed at a randomly selected position on an all-zero image with a resolution of $56 \times 56$ pixels. The position must ensure completeness of the interference and minimize overlap with target digits. A total of 20,000 images are generated and labeled as 6.

5. Shuffling: Finally, a total of 40,000 images from above datasets are shuffled to create the training dataset.

**(2) Loss function**

We designed separate loss functions for the target and the interference. For the target loss function, the index of the classification region with the highest intensity on the output plane is required to match the label. To achieve that, the difference between the optical field intensity among the six classification regions and the one-hot encoding of the target label is computed using mean squared error (MSE) loss function.

Meanwhile, to ensure the energy is concentrated as much as possible in the classification regions, we calculated the ratio of the maximum energy among the six classification regions to the total energy sum of the six regions, and make it close to 1 during the optimization process. The mathematical expression is as follows:

$$Loss_{target} = \beta_1 \times MSE(O^+, G^+) + \beta_2 \times (1 - E) \tag{S4}$$

For the interference loss function, we employed Pearson Correlation Coefficient (PCC) to make interference output field distribution like random noise outside the classification regions. The components of this loss function are discussed specifically in the main text. The mathematical expression is as follows:

$$Loss_{inter} = a_1 \times PCC(O, I) + a_2 \times (1 - PCC(O, O_{DET})) + a_3 \times PCC(O, O_{SFT}) \tag{S5}$$

Here, O and I means optical field distribution in output plane and input plane separately, $O_{DET}$ means optical field avoiding all classification regions, $O_{SFT}$ means optical field shifts several pixels.

Each training batch consists of 8 images, which include both targets labeled as 0 to 5 and interference labeled as 6. Different calculation methods for the loss function were selected according to the labels, and all the loss values in one batch were summed up.

$$Loss = Loss_{target} + Loss_{inter} \tag{S6}$$

The stochastic gradient descent (SGD) algorithm and the Adam optimizer was employed to update the phase value of each layer in the network. During the optimization process, the loss gradually decreases, and ultimately the expected output optical field effect is achieved.

**Supplementary Note 3: Analysis of robust training**

AI D$^2$NN operates at the working frequency of 0.85 THz. To ensure the diffractive propagation of the optical field in free space, it's not recommended for the size of each diffractive neuron more than 0.5 λ. In the main text, we compare the training accuracies of diffractive neural networks with neuron sizes of 100 μm (0.28 λ) and 200 μm (0.56 λ), and confirm that the network with 100 μm neurons exhibits more excellent performance compared to 200 μm counterpart. Considering fabrication constraints and alignment difficulty, we selected the 100 μm neuron size and a 2-layer configuration for subsequent experiments.

Other model parameters are set as follows: the distance from the input plane to the first diffractive layer is set to 1 cm, the spacing between diffractive layers is 0.5 cm, and the distance from the last diffractive layer to the output plane is 1 cm. However, in practical experiments, experimental and fabrication errors often lead to significant degradation in model performance. Therefore, robustness training is necessary, which involves introducing random errors such as transverse shifts, rotational misalignments, and displacement errors along the z-axis during the training process. Table S1 compares the combined effects of multiple experimental errors on network performance, which are characterized by the test accuracies. It can be seen that as the random errors increase, the performance gap between the network without robustness training and one with robustness training widens. It is evident that robustness training is crucial for the network to resist various errors and maintain good inference performance.

**Table S1 The combined effects of experimental errors on the performance of network**

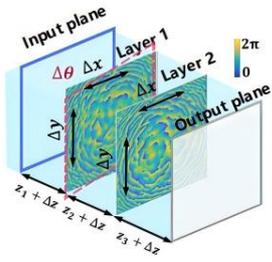

| | Δx (pixel) | Δy (pixel) | Δθ (°) | Δd (mm) | w/o robustness test accuracy (%) | w/ robustness test accuracy (%) |
|---|---|---|---|---|---|---|
| case1 | 0 | 0 | 0 | 0 | 89.3 | 89.3 |
| case2 | 1 | 0 | 1 | 0.1 | 84.2 | 88.7 |
| case3 | 0 | 1 | 1 | 0.1 | 84.3 | 88.6 |
| case4 | 1 | 1 | 1 | 0.1 | 82.1 | 87.4 |
| case5 | 2 | 2 | 1 | 0.1 | 71.2 | 83.6 |
| case6 | 2 | 2 | 1 | 0.2 | 67.6 | 81.5 |

**Supplementary Note 4: Detailed experimental setup and fabricated samples**

The experimental setup, including the alignment of the input mask with the metasurface, is shown in Fig. S1. To ensure precise alignment and spacing between the input mask and two metasurfaces, a high-precision 3D print holder was employed. During experiments, the stainless steel mask with an etched input pattern is positioned 10 mm in front of the metasurface to create a predefined spatial intensity distribution. The two metasurfaces are firmly inserted into the holder, maintaining a 5mm spacing. The crystal ZnTe is positioned 10 mm behind the metasurface to capture the transmitted signals. The pump beam passes through the mask and metasurfaces, with incidence on the crystal ZnTe and coincidence with the probe beam. The probe beam is reflected from the left surface of the crystal ZnTe, whose polarization state is modulated by the THz electric field according to the Pockels effect. The probe beam carrying terahertz information exits the crystal ZnTe, and passes through a quarter-wave plate and a Wollaston prism. The CCD acquires the THz time-domain pulse signal via scanning sampling, which is then converted into the THz frequency-domain spectrum through

Fourier transformation.

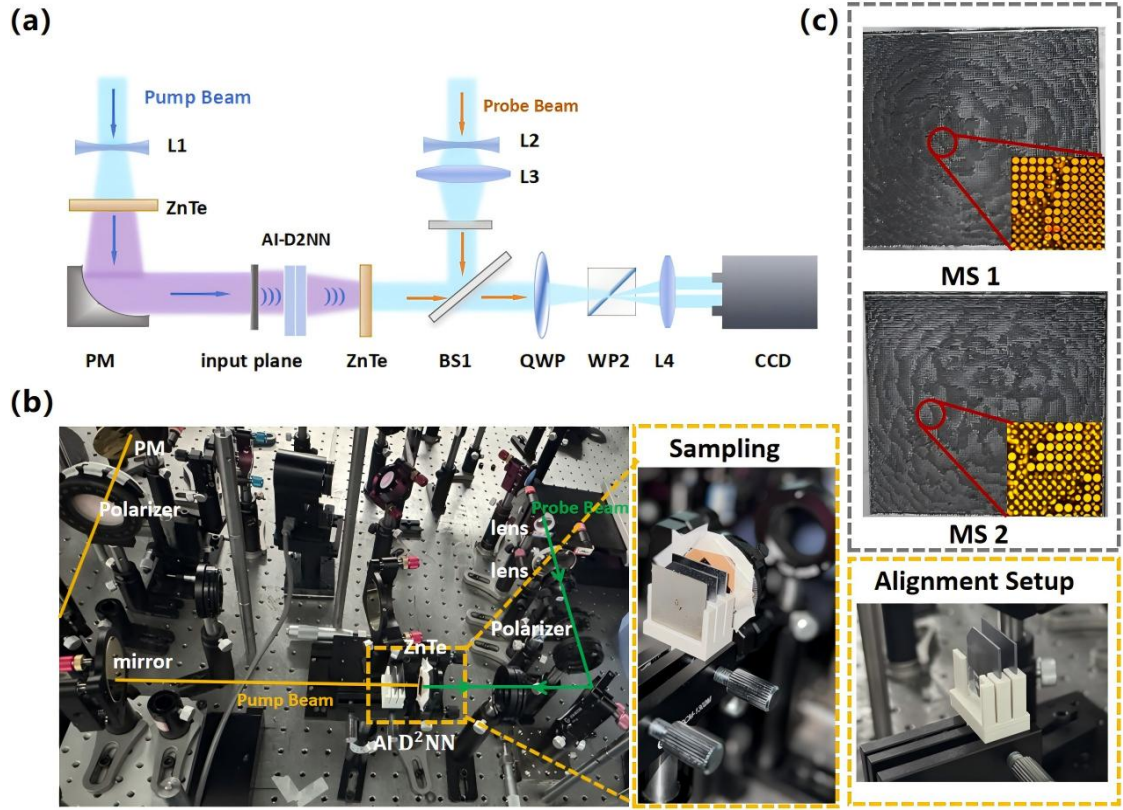

**Fig. S1 Experimental setup and fabricated samples.** (a) The schematic of THz experimental platform (b) Photograph of the experimental setup (c) Optical images of the fabricated matasurface. Insets in (c) show the zoom-in microscopic images of meta-atoms.

**Supplementary Note 5: The structural parameters of cylindrical rods**

For ease of fabrication, we selected 16 types of structures whose amplitude coefficients close to 1 and phase responses nearest integer multiples of Π/8, and assembled these structures according to the neuron values of the trained diffractive neural network. The parameters of selected cylindrical rods are as follows:

**Table S2 The structural parameters of cylindrical rods**

| label | 1 | 2 | 3 | 4 | 5 | 6 | 7 | 8 |
|---|---|---|---|---|---|---|---|---|
| Radius($\mu m$) | 15 | 20 | 23 | 25 | 28 | 30 | 32 | 33 |
| label | 9 | 10 | 11 | 12 | 13 | 14 | 15 | 16 |
| Radius($\mu m$) | 35 | 36 | 37 | 38 | 39 | 41 | 42 | 43 |

The phase profile corresponding to the structure labeled as 1 is -7/8 Π, structure labeled as 2 is -6/8 Π.... structure labeled as 16 is Π.

**Supplementary Note 6: Energy distribution for the AI D²NN**

Analyzing the electric field distribution at the output plane is crucial for assessing the network's capability to recognize targets 0-5 in multi-object scenes. In addition, other evaluation metrics such as energy efficiency E, normalized energy contrast $\triangle$E and signal-to-noise ratio SNR, must be considered to ensure that the classification results are distinct and reliably represented.

First, the energy distribution across 6 classification regions is calculated as follows:

$$\begin{bmatrix} P_0 = \sum i_1 \leq i \leq i_2 \sum j_1 \leq j \leq j_2 \mid E_{(i,j)} \mid^2 \\ P_1 = \sum i_3 \leq i \leq i_4 \sum j_3 \leq j \leq j_4 \mid E_{(i,j)} \mid^2 \\ P_2 = \sum i_5 \leq i \leq i_6 \sum j_5 \leq j \leq j_6 \mid E_{(i,j)} \mid^2 \\ P_3 = \sum i_7 \leq i \leq i_8 \sum j_7 \leq j \leq j_8 \mid E_{(i,j)} \mid^2 \\ P_4 = \sum i_9 \leq i \leq i_{10} \sum j_9 \leq j \leq j_{10} \mid E_{(i,j)} \mid^2 \\ P_5 = \sum i_{11} \leq i \leq i_{12} \sum j_{11} \leq j \leq j_{12} \mid E_{(i,j)} \mid^2 \end{bmatrix} \tag{S7}$$

In each row, i and j denote the starting and ending positions of the column and row coordinates in the classification regions, respectively. In this work, a classification region occupies 7×7 pixels. Six classification regions totally occupy 294 pixels. Since the classifier is designed for six-class classification, the total energy in all recognition regions is given by:

$$P_{\text{total}} = P_0 + P_1 + P_2 + P_3 + P_4 + P_5 \tag{S8}$$

$$P_{out} = \sum_{i=1}^{100} \sum_{j=1}^{100} \mid E_{\text{output}(i,j)} \mid^2 \tag{S9}$$

$$P_{input} = \sum_{i=1}^{100} \sum_{j=1}^{100} \mid E_{\text{input}(i,j)} \mid^2 \tag{S10}$$

The energy efficiency for the maximum energy among six regions is calculated as:

$$E = \frac{P_{\max}}{P_{\text{total}}} \tag{S11}$$

The normalized energy contrast △E represents the ratio of the difference between the highest and second-highest energies across the six detection regions to the highest energy, which can be expressed as:

$$\Delta E = \frac{P_{max} - P_{second}}{P_{max}} \tag{S12}$$

The signal-to-noise ratio SNR here is defined as the ratio of the maximum average intensity of the classification regions to the average intensity of all other areas on the output plane excluding the six detection regions, the mathematical expression is as follows:

$$SNR = \frac{P_{max}/49}{(P_{out} - P_{total})/(10000-49)} \tag{S13}$$

**Supplementary Note 7: Trained network's classification results of different cases**

This section provides further details about the experimental results, specifically supplementing the recognition performance analysis of the trained AI $D^2NN$ for handwritten digits classification in multi-object scenarios.

First, the optical field distributions of the network recognizing a single digit (0–5) or an interfering object are illustrated in Fig. S2.

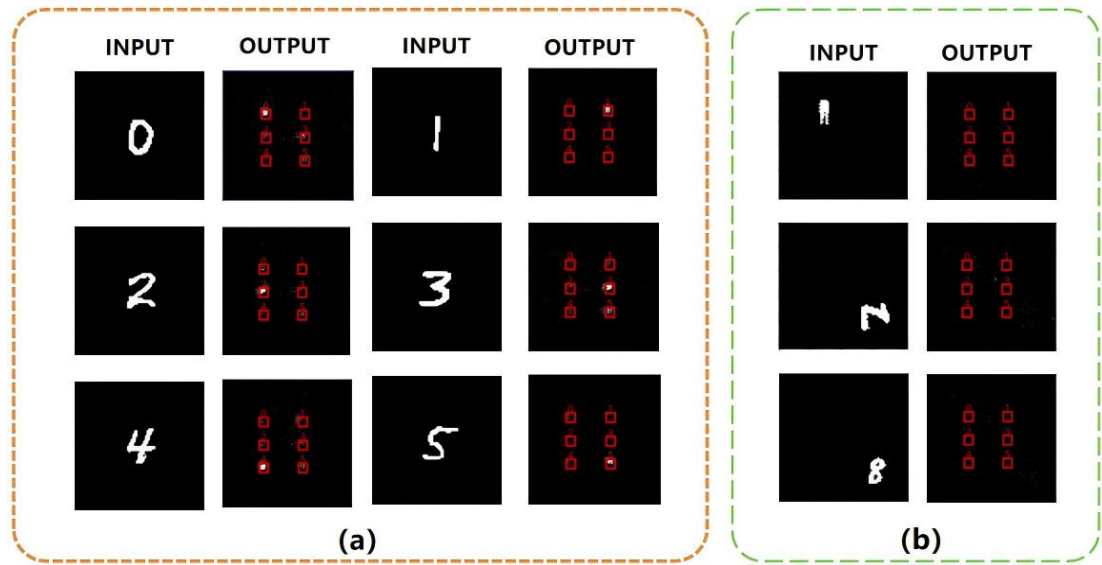

**Fig. S2 Trained network's classification results of different objects.** (a) Simulation results of classifying single target (b) Simulation results of classifying single interference

Next, the recognition accuracies and confusion matrices of the AI $D^2NN$ for the intra-class, inter-class, and dynamic interference are presented in Fig. S3.

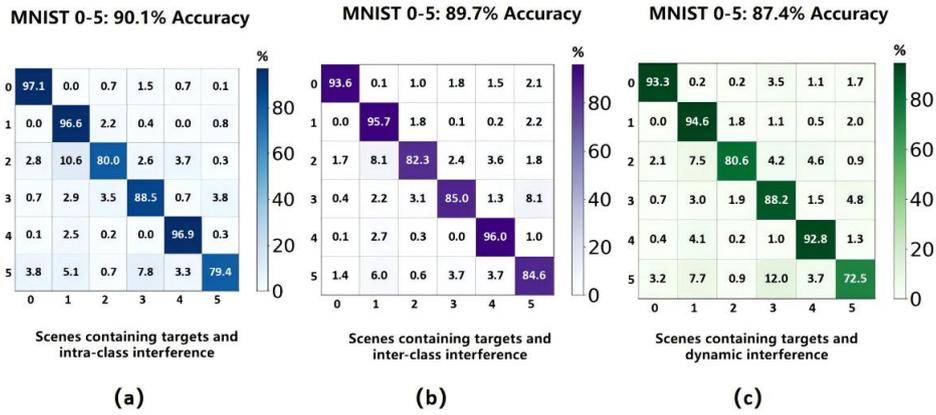

**Fig. S3 The recognition accuracies and confusion matrices of the AI D²NN for the three types of interference.** (a) Intra-class interference (b) Inter-class interference (c) Dynamic interference

Besides, the AI D²NN demonstrates the capability to recognize handwritten digits 0-5 in multi-object scenes. Subsequently, we randomly tested scenarios containing intra-class interference, inter-class interference, and dynamic interference. The results of network classification performance for digits 0-5 are shown in Fig. S4.

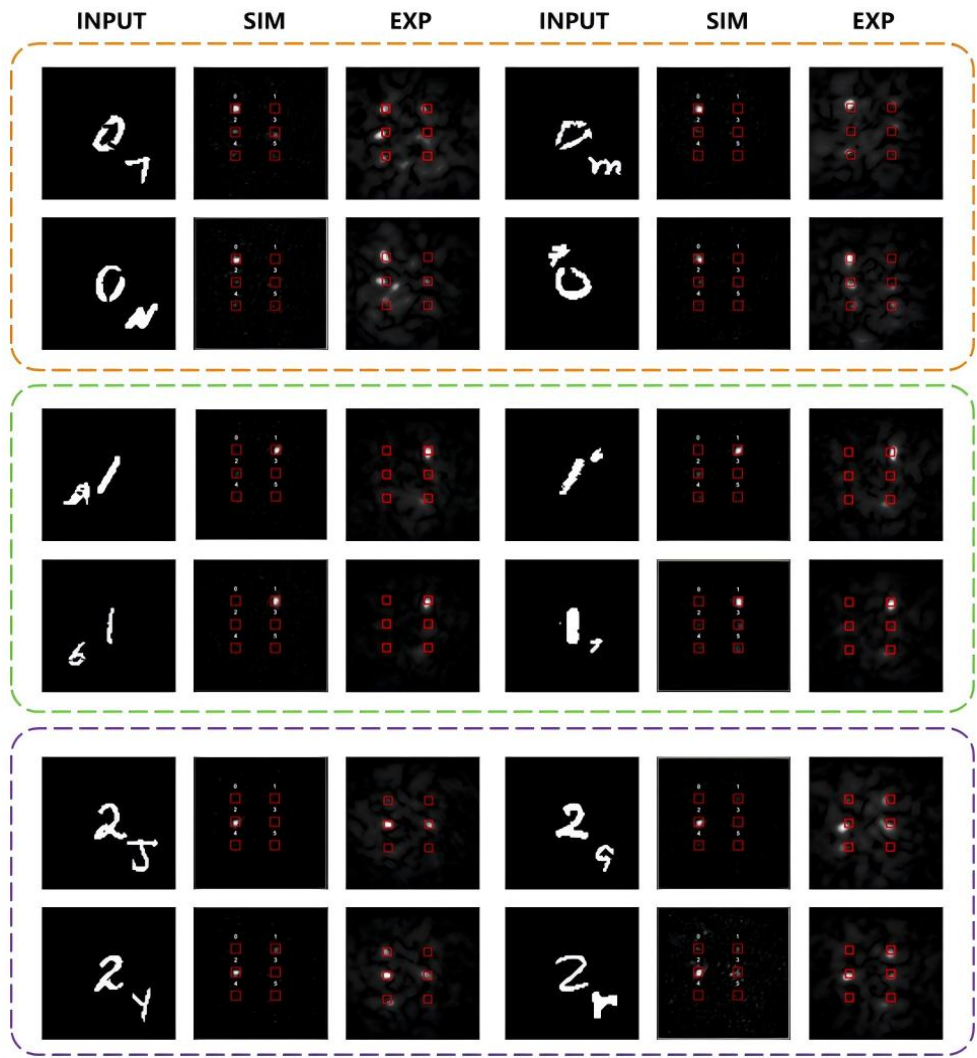

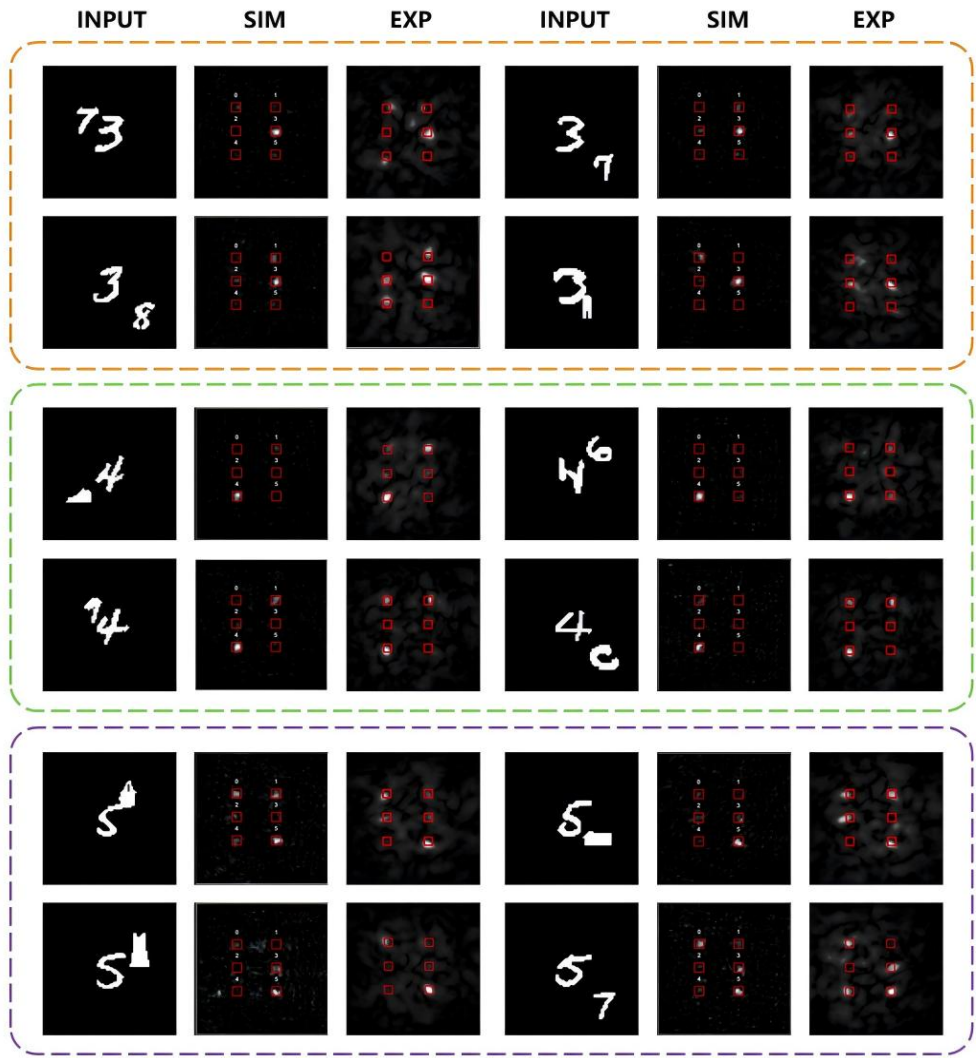

**Fig. S4 Energy distribution of numerical and experimental results for the classification of digits 0-5 in multi-object scenes**

Optical neural networks can perform computations at the speed of light. Therefore, the AI D²NN is capable of recognizing handwritten digits in scenes containing dynamic interference. The simulation results are presented in **Fig. S5.**

The concept of transforming the optical fields of non-target objects into background noise through ONN is inherently extendable, for example, even as the number and categories of interference increase, the network still demonstrates excellent capability in classifying the primary target in multi-object scenarios. In theory, more complex scenarios for target recognition can be achieved by employing more diverse and extensive datasets, more rigorous loss functions and deeper network models during training phase. Therefore, we constructed a 5-layer AI D²NN to adapt to scenarios featuring a single target and more than two interfering objects, where the positions and sizes of all interfering objects vary flexibly. The simulation results of classifying 0-5 when containing 2 interfering objects are shown in Fig. S6. The simulation results of classifying 0-5 when scenes containing 2 moving interfering objects are presented in Fig. S7. The simulation results of classifying 0-5 when scenes containing 3 moving interfering objects are illustrated in Fig. S8.

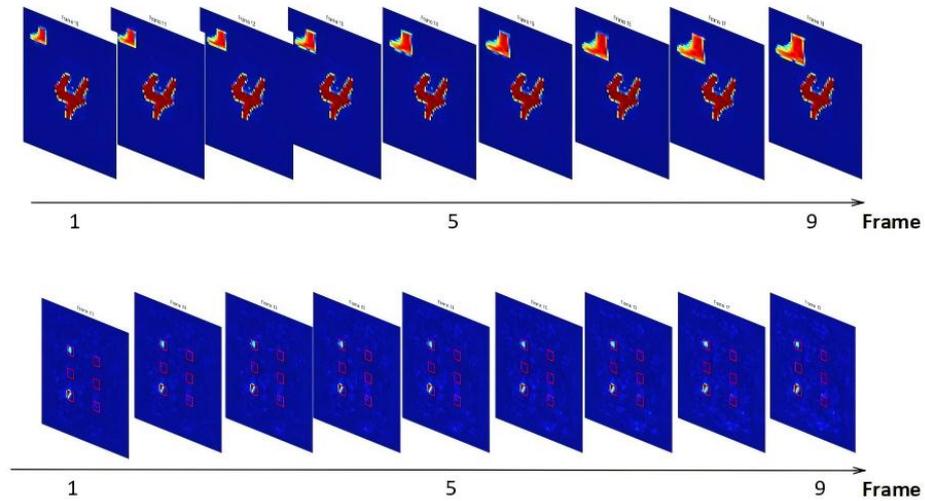

**Fig. S5 Energy distribution of numerical results for the classification of digits 0-5 in scenes containing dynamic interference.**

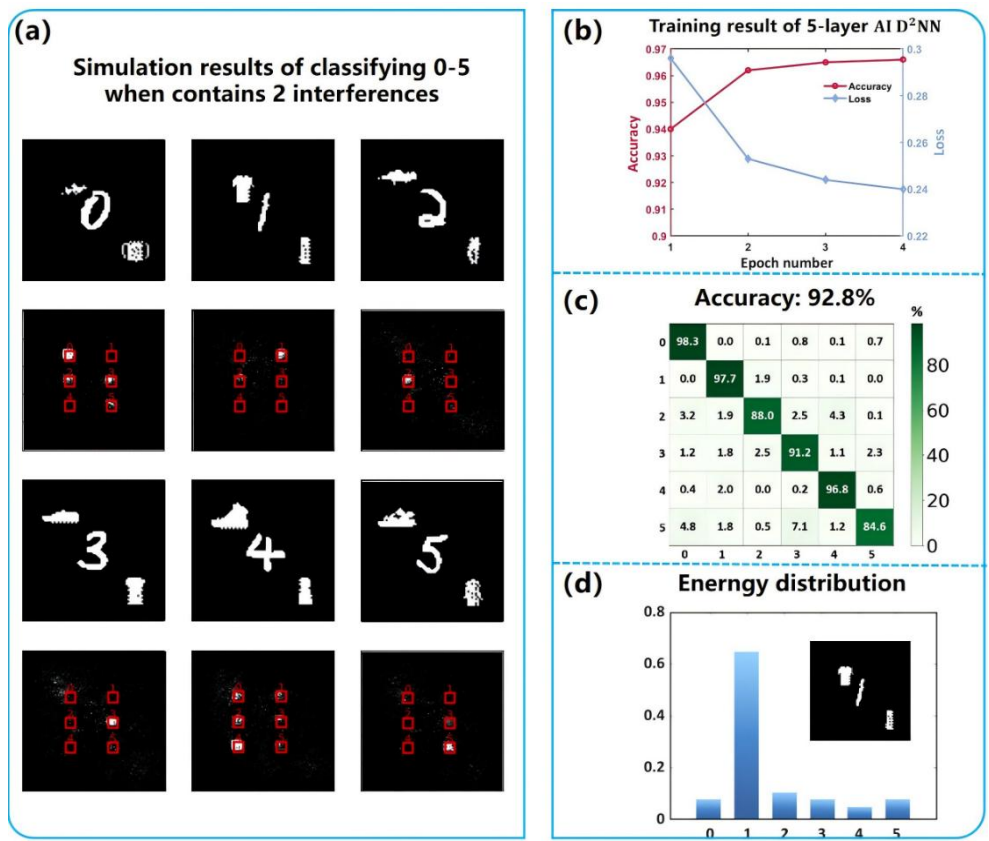

**Fig. S6 Simulation results of classifying 0-5 when scenes containing 2 interferences.** (a) The output plane's energy distribution (b) Training result (c) Confusion matrix of test dataset and test accuracy (d) Specific energy distribution when scenes containing digit 1, trousers and T-shirt

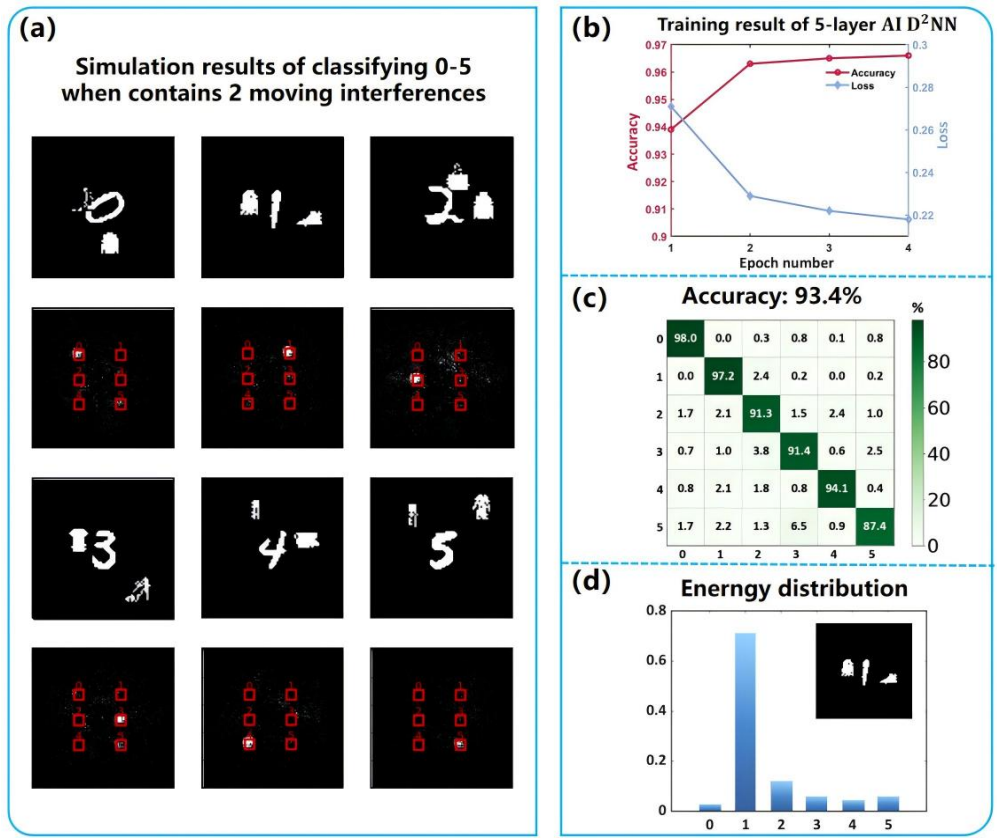

**Fig. S7 Simulation results of classifying 0-5 when scenes containing 2 moving interferences.** (a) The output plane's energy distribution (b)Training result (c) Confusion matrix of test dataset (d) Specific energy distribution when scenes containing digit 1, coat and shoes

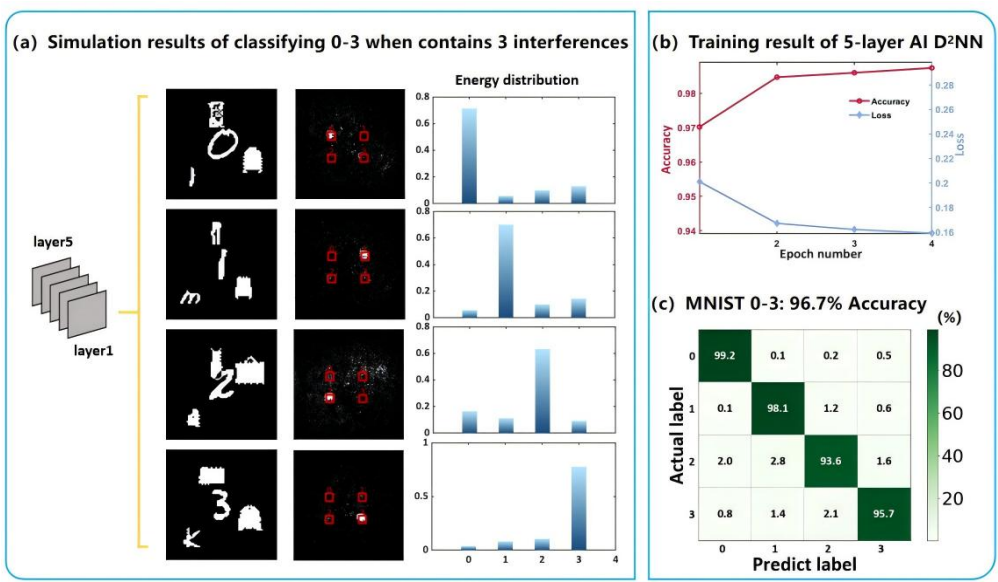

**Fig. S8 Simulation results of classifying 0-3 scenes when containing 3 moving interferences.** (a) The output plane's energy distribution (b) Training result (c) Confusion matrix of test dataset